\documentclass[10pt,conference]{IEEEtran}

\usepackage{cite}
\usepackage[linesnumbered,ruled]{algorithm2e}
\usepackage{url} 

\usepackage{balance}
\usepackage{graphicx}
\usepackage{textcomp}
\usepackage{xcolor}
\usepackage{comment}
\usepackage{booktabs}
\usepackage{multirow}
\usepackage{xspace}
\usepackage{tcolorbox}
\usepackage[dvipsnames]{xcolor}

\usepackage{amsmath}
\usepackage{amssymb}
\usepackage{mathtools}
\usepackage{amsthm}

\usepackage{array}

\usepackage{subcaption}

\SetKwComment{Comment}{/* }{ */}

\newcommand{\PreserveBackslash}[1]{\let\temp=\\#1\let\\=\temp}
\newcolumntype{C}[1]{>{\PreserveBackslash\centering}p{#1}}
\newcolumntype{R}[1]{>{\PreserveBackslash\raggedleft}p{#1}}
\newcolumntype{L}[1]{>{\PreserveBackslash\raggedright}p{#1}}

\newif\ifshowtodos
\showtodostrue

\def\BibTeX{{\rm B\kern-.05em{\sc i\kern-.025em b}\kern-.08em
    T\kern-.1667em\lower.7ex\hbox{E}\kern-.125emX}}
    
\newcommand{\pa}[1]{\noindent\textbf{#1}}
\newcommand{\tool}{\textit{ADPerf}}

\newcommand{\rqboxc}[1]{\begin{tcolorbox}[left=4pt,right=4pt,top=4pt,bottom=4pt,colback=gray!5,colframe=gray!40!black,before skip=4pt,after skip=4pt]#1\end{tcolorbox}}

\def\BibTeX{{\rm B\kern-.05em{\sc i\kern-.025em b}\kern-.08em
    T\kern-.1667em\lower.7ex\hbox{E}\kern-.125emX}}

\begin{document}

\title{\tool: Investigating and Testing Performance in Autonomous Driving Systems}

\author{\IEEEauthorblockN{Tri Minh-Triet Pham}
\IEEEauthorblockA{\textit{O-RISA Lab} \\
\textit{Concordia University}\\
Montreal, Quebec, Canada \\
p\_triet@encs.concordia.ca}
\and
\IEEEauthorblockN{Diego Elias Costa}
\IEEEauthorblockA{\textit{REALISE Lab} \\
\textit{Concordia University}\\
Montreal, Quebec, Canada \\
diego.costa@concordia.ca}
\and
\IEEEauthorblockN{Weiyi Shang}
\IEEEauthorblockA{\textit{SENSE Lab} \\
\textit{University of Waterloo}\\
Waterloo, Ontario, Canada \\
wshang@uwaterloo.ca}
\and
\IEEEauthorblockN{Jinqiu Yang}
\IEEEauthorblockA{\textit{O-RISA Lab} \\
\textit{Concordia University}\\
Montreal, Quebec, Canada \\
jinqiu.yang@concordia.ca}
}

\maketitle

\begin{abstract}
Obstacle detection is crucial to the operation of autonomous driving systems, which rely on multiple sensors, such as cameras and LiDARs, combined with code logic and deep learning models to detect obstacles for time-sensitive decisions. Consequently, obstacle detection latency is critical to the safety and effectiveness of autonomous driving systems.
However, the latency of the obstacle detection module and its resilience to various changes in the LiDAR point cloud data are not yet fully understood.
In this work, we present the first comprehensive investigation on measuring and modeling the performance of the obstacle detection modules in two industry-grade autonomous driving systems, i.e., Apollo and Autoware.
Learning from this investigation, we introduce \tool, a tool that aims to generate realistic point cloud data test cases that can expose increased detection latency. 
Increasing latency decreases the availability of the detected obstacles and stresses the capabilities of subsequent modules in autonomous driving systems, i.e., the modules may be negatively impacted by the increased latency in obstacle detection.

We applied \tool{} to stress-test the performance of widely used 3D obstacle detection modules in autonomous driving systems, as well as the propagation of such tests on trajectory prediction modules.
Our evaluation highlights the need to conduct performance testing of obstacle detection components, especially 3D obstacle detection, as they can be a major bottleneck to increased latency of the autonomous driving system. Such an adverse outcome will also further propagate to other modules, reducing the overall reliability of autonomous driving systems.
\end{abstract}

\begin{IEEEkeywords}
Autonomous vehicles, Lidar, Software testing, System testing, Simulation, Performance evaluation
\end{IEEEkeywords}

\section{Introduction}
Autonomous driving offers the potential for safer, more reliable, and more efficient transportation by reducing human error and enabling rapid, complex decision-making.
At the core of any autonomous driving system (ADS) is the perception module, which is safety-critical.
Leveraging algorithms and deep learning (DL) models, this module provides the ADS with real-time interpretation of its environment, allowing it to navigate safely using data from a variety of sensors, including LiDAR and cameras.
As part of perception, obstacle detection performs the crucial task of identifying and localizing surrounding vehicles, pedestrians, etc.
As a result, the accuracy of these detectors is critical, as any failure to correctly identify an obstacle could lead to critical safety issues, such as collisions.
As such, a large body of research has been devoted to testing these components \cite{Tang2022ASO}. This includes testing detectors' robustness towards noises and weather corruptions \cite{Dong2023BenchmarkingRO, metamorphic_lidar}, adversarial perturbations \cite{ShapeShifter, invisibleobject, gan_attack,liu_multiview, Xu2021AdversarialAA, Hau2021ObjectRA}, etc. on the sensor data. However, their focus has been on accuracy and robustness while the performance is understudied \cite{ml_testing_survey, Tang2022ASO}. 

The performance of obstacle detection is a critical yet underexplored area, directly influencing safety and reliability. Often referred to as availability or efficiency, this performance reflects the system’s detection speed. High latency can impair decision-making, and delayed obstacle detection is as hazardous as an undetected one, leaving the ADS with insufficient time to respond.
Addressing this challenge is vital for two reasons. First, the accuracy and efficiency of detection are highly sensitive to complex, unpredictable real-world workloads, which differ from the controlled datasets typically used in evaluations. Current workloads often fail to reveal performance issues, leaving potential vulnerabilities that could lead to unforeseen failures during deployment.
However, existing performance testing research has primarily concentrated on traditional software systems, often overlooking the evaluation of complex AI system performance \cite{Jangali2023AutomatedGA}.
While recent works have analyzed the impact of adversarial perturbations on 3D \cite{slowlidar} and 2D \cite{Shapira2022PhantomSE, Chen2023OverloadLA, Ma2023SlowTrackIT,10650435} obstacle detection, the performance of obstacle detection in particular and perception in general under normal circumstances is not yet fully understood. Furthermore, there remains a gap in availability-based robustness testing of obstacle detection.

To address these concerns, in this work, we present the first performance study analyzing and modeling two representative AD systems, i.e., Apollo \cite{apollo} and Autoware \cite{autoware}. First, we model the entire MSF obstacle detection module, which includes 2D and 3D obstacle detection and MSF, in the two AD systems using Queuing Networks to construct a Queuing Petri-Net (QPN) in QPME. Then, we exercise the obstacle detection modules using diverse driving scenarios (nuScenes \cite{nuscenes} for Apollo and AWSIM \cite{awsim} for Autoware) to compute the service rate of the components in the QPN. Afterwards, we use the measurement to configure the built performance models. Finally, we simulate them in QPME to identify throughput and bottleneck issues. Our simulation results reveal that 3D obstacle detection represents a critical bottleneck for detection, significantly hindering overall system performance.



Motivated by the modeling and simulation results, we introduce \tool, a novel tool designed to generate new performance tests by leveraging existing test scenarios  (i.e., PCD) that test 3D obstacle detection. Furthermore, \tool{} records latency and computes data availability, and then evaluates the impacts of frame dropping on trajectory prediction. The outcome of \tool{} can be used to estimate frame availability, identifying which frames are dropped due to high latency. We conduct an evaluation using two widely-used LiDAR obstacle detection models in Level 4 AD systems: PointPillars in Apollo and CenterPoint in Autoware \cite{autoware}, using the nuScenes dataset. 
Our evaluation results show that \tool{} amplifies the latency of 3D obstacle detection, increasing the average latency by up to 11.5ms. This increase in latency leads to a 9\% rise in the percentage of dropped frames. The resulting decline in 3D obstacle detection performance cascades downstream, causing significant deviations in trajectory prediction and underscoring the critical impact of latency on system reliability.

Our work makes the following contributions:

\begin{itemize}

    \item We present the first study analyzing and modeling the performance of two Autonomous Driving Systems (Apollo and Autoware). Our analysis reveals the performance bottlenecks in the two ADSs.
    \item We propose a testing approach (\tool) that automatically generates latency-stressing tests for 3D obstacle detection.
    \item We applied \tool{} to examine the performance of Apollo and Autoware through three experiments. First, we evaluate the robustness of 3D obstacle detection, i.e., PointPillar and CenterPoint, and measure the change to latency and availability. Second, we evaluate the robustness of trajectory prediction, i.e., Trajectron++, to the unavailability of 3D obstacle detection outputs. Last, we experiment with the impact of the latencies of the 3D obstacle detection on the subsequent trajectory prediction module through a simulated evaluation. 
\end{itemize}

\pa{Paper organization.} Section~\ref{sec:background} presents the background and related work.
Section~\ref{sec:setup} introduces our experiment setup.
Section~\ref{sec:modeling} presents the first ADS performance modeling study on both Apollo and Autoware.
Section~\ref{sec:adf} introduces \tool, an automated performance test
generation approach for ADS, followed by Section \ref{sec:adperf_eval}, a comprehensive evaluation of \tool.
Section~\ref{sec:threats} discusses the threats to the validity, and Section~\ref{sec:conclusion} concludes the paper.

\section{Background and Related Work}
\label{sec:background}
In this section, we present the background of ADSs, with a focus on 3D obstacle detection and trajectory prediction. Then, we describe previous work that tests or attacks these modules in ADSs. Moreover, we discuss performance testing and attacking in the ADS obstacle detection context, as well as general performance testing and modeling.
\begin{figure}[]
  \centering
  \includegraphics[width=\linewidth]{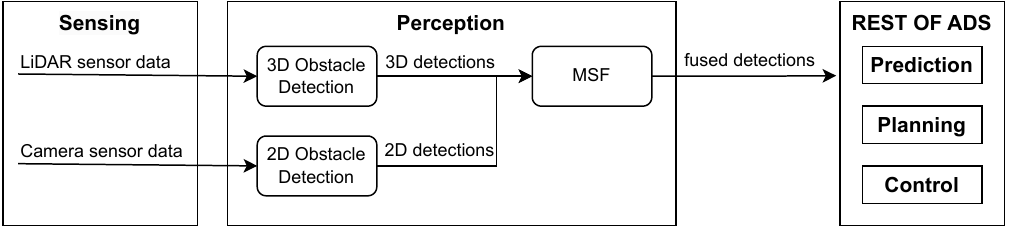}
  \caption{The data flow between the modules (with a focus on sensing and perception components) in a general ADS. Arrows indicate data channels.
  }
  \label{fig:apollo_sensing_perception}
\end{figure}

At a high level, ADSs are composed of several functional modules, i.e., sensing, perception, prediction, planning, and control~\cite{Tang2022ASO} as shown in Figure~\ref{fig:apollo_sensing_perception}. 
The modules and their components communicate via messages/frames. The underlying middleware frameworks typically used are the Robot Operating System (ROS2)~\cite{Thomas2014} (for Autoware) or CyberRT~\cite{cyberrt} (for Apollo). 
The communication is organized using a publisher-subscriber model for topics/channels.
In Apollo, obstacle detection is a key part of perception, which consists of three components: 3D obstacle detection, 2D obstacle detection, and multi-sensor fusion (MSF). The output of MSF would be propagated to the rest of the ADS.
In Apollo's MSF, LiDAR is the main sensor \cite{apollo_main_sensor_sample}, meaning that the MSF process waits for a LiDAR scan before integrating data from other sensors \cite{apollo_msf_component, apollo_prob_fusion}, such as cameras and radar. Consequently, the throughput of MSF is limited by the performance of 3D obstacle detection. Similarly, Autoware only performs 3D obstacle detection; hence, 3D obstacle detection throughput is obstacle detection throughput.

\subsection{Detection robustness of 3D obstacle detection}
ADS-mounted LiDAR scans the environment at regular intervals, producing point cloud data (PCD) where each point represents the 3D location of a reflected laser beam. 3D obstacle detection identifies obstacles in the PCD by fitting 3D bounding boxes around them.

The detection process is sensitive to various tests and attacks, such as the addition of random points outside the region of interest (ROI) in 3D obstacle detection~\cite{metamorphic_lidar}, as well as weather and sensor-based corruptions to the PCD~\cite{yu2022benchmarking, Gao2023BenchmarkingRO, leimalidarrobustnessbenchmark, LiRtest, Dong2023BenchmarkingRO, pham2024evaluatingrobustnesslidarbased3d}.
Instead of modifying the sensed data,~\cite{pham2024perceptionguidedfuzzingsimulatedscenariobased} fuzzed the simulated driving scenes to evaluate the robustness of the perception system in ADS. 
Most of these works focus on reducing the number of obstacles detected, thereby decreasing latency. In contrast, \tool{} is designed to generate additional detections, which has the potential to increase latency.

Previous 3D obstacle detection attacks have included 
spoofing obstacles~\cite{adversarial_lidar_attack, Illusion_and_Dazzle}, perturbing vulnerable regions in PCD \cite{Xu2021AdversarialAA,Hau2021ObjectRA, ZHENG2023109825,adversarial_locations,Zhu2021AdversarialAA}, 
conducting occlusion-guided optimization-based black-box attacks~\cite{lidar_occulsion,WANG202127}, leveraging multiple sensors for GAN-based attacks~\cite{gan_attack,liu_multiview}, generating adversarial obstacles~\cite{Yang2021RobustRP,Abdelfattah,pmlr-v164-tu22a,Tu_2020_CVPR,invisibleobject}, or spoofing the ADS's trajectory~\cite{li2021fooling}.
Among these works, some increase the number of detected obstacles through physical attacks \cite{adversarial_lidar_attack} or adversarial patches \cite{Illusion_and_Dazzle, pmlr-v164-tu22a}. However, these methods typically result in only a very small increase in detection, as they are difficult to execute and require object-specific patches. In contrast, our approach does not involve attacking 3D obstacle detection. Instead, we assess performance robustness by significantly increasing the number of detections to induce higher latency.

\subsection{Trajectory prediction tests and attacks}
Trajectory prediction forecasts the future trajectories of detected obstacles over the next few seconds based on their historical positions, headings, velocities, and other factors. Previous works testing this module have involved making obstacles appear or disappear for robustness testing~\cite{metamorphic_prediction} or generating safety-critical scenarios using LiDAR for system evaluation~\cite{advsim}. Attacks have included adversarial perturbations to the historical trajectory~\cite{Zhang_2022_CVPR} or detection bounding boxes~\cite{attack_tracking}, as well as attention-guided perturbations aimed at disrupting pedestrian-based collision avoidance~\cite{SAADATNEJAD2022103705}. In contrast to existing works that test or attack trajectory prediction through perturbations to historical data or safety-critical scenarios, we approach trajectory prediction from the perspective of data availability, specifically examining the robustness of trajectory prediction when faced with data unavailability.

\subsection{Performance testing}
\label{sec:related_perf}
One key efficiency metric of a machine learning system is the prediction or inference speed~\cite{ml_testing_survey}.
In real systems, the prediction speed can become more important than model accuracy, particularly as data and system complexity grow~\cite{BaezaYates2017QualityefficiencyTI}. In this work, we refer to efficiency as detection latency. Software performance testing assesses a system's performance under different conditions to ensure it meets requirements and operates efficiently at both the system and component levels.
Performance testing measures a system’s performance (e.g., response time, utilization, and throughput) under a specific workload~\cite{7123673,9813593}.

\pa{Model-based Performance Analysis.}
Model-based performance analysis involves constructing abstract models to predict and analyze the performance of software systems~\cite{Balsamo2004ModelbasedPP}. These models describe the system’s components, communication, and resource utilization, facilitating simulation and analysis~\cite{Becker2007ModelBasedPP, Becker2009ThePC}. Such models can be represented in various forms, each capturing different characteristics of the software, such as queueing networks (QN)~\cite{kleinrock1974queueing, kant1992introduction} and queueing Petri nets (QPN)~\cite{Bause1993QNP, Bause1993QueueingPN, Baccelli1994AnnotatedBO, Bause1997IntegratingSA}. These models offer an efficient way to understand the performance of software systems and identify potential bottlenecks under varying workloads. 
In this work, we model and simulate an ADS, enabling its performance analysis under different workloads.

\pa{Latency Robustness of Obstacle Detection.} Most previous work has focused on the obstacle detection correctness instead of the performance.
Of those, two studies propose adversarial attacks that generate perturbations to maximize the number of candidate proposals ($K$) in the post-processing stage, exploiting the $O(K^2)$ worst-case complexity of this step to maximize the latency for 2D obstacle detection~\cite{Shapira2022PhantomSE, Chen2023OverloadLA, Ma2023SlowTrackIT,10650435} and 3D obstacle detection~\cite{slowlidar}. 
While both ADPerf and SlowLidar \cite{slowlidar} perturb the PCD to increase latency, ADPerf adopts a more realistic approach. Unlike SlowLidar, which relies on an adversarial attack requiring multiple optimization steps and alters 3,000 points per iteration, ADPerf is a black-box testing method that introduces fewer than 670 modifications in a single step without depending on internal post-processing stages. Making fewer changes preserves the semantic integrity of the frame and aligns better with potential real-world or physical perturbations. 
We aim to observe the effect of noises and realistic modifications to the PCD on the latency of 3D obstacle detection.

\section{Study Overview and Experimental Setup}
\label{sec:setup}
\begin{figure}[]
  \centering
  \includegraphics[width=\linewidth]{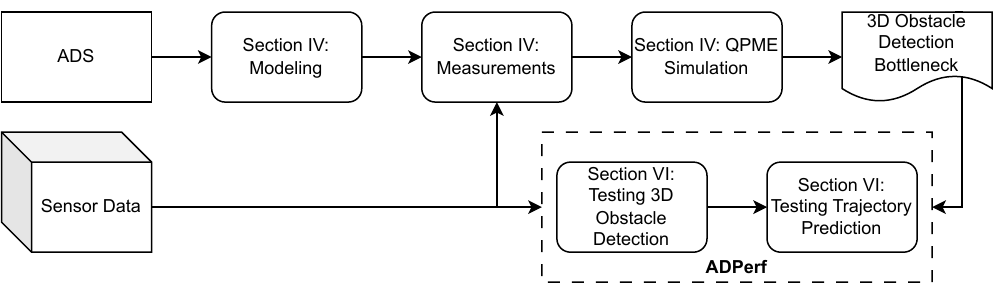}
  \caption{Overview of our experiment}
  \label{fig:overview_of_experiment}
\end{figure}

\pa{Study Overview.} The experiment consists of two stages. First, we study the performance of the MSF obstacle detection module in two ADS to identify the bottleneck, focusing on 3D obstacle detection. Second, we assess the robustness of 3D obstacle detection to latency-increasing tests and evaluate its cascading impact on the robustness of trajectory prediction. To confirm our findings, we simulate a scenario to demonstrate the effects of increased 3D obstacle detection latency on the entire ADS. We summarize the experiment in Figure \ref{fig:overview_of_experiment}.

\pa{Experiment Setup.} Our experiments were conducted on an Ubuntu 20.04.3 LTS system with an AMD Ryzen 16-core CPU, 256 GiB of RAM, and an NVIDIA GeForce RTX 3090 GPU. 

We set up two ADSs for the modeling experiment (Section \ref{sec:modeling}), i.e., Autoware (RobotecAI's SS2 stable version) \cite{autoware_robotec} and Apollo V7 \cite{apollo}, and along with them, two sources of input, i.e., V1.2.0 SS2 \cite{awsim} and nuScenes \cite{nuscenes}. While AWSIM is the best digital twin simulator for Autoware, we also chose nuScenes for our experiment because it is a popular dataset for ADS testing, which includes driving scenarios that require consecutive sensor inputs. Additionally, the tests for \tool{} necessitate obstacle annotations for those consecutive inputs, which datasets like KITTI~\cite{Geiger2012CVPR} do not provide.
nuScenes is a large-scale dataset featuring a wide range of complex driving scenarios for autonomous driving. It includes 1,000 20-second scenes collected in Boston and Singapore. The LiDAR sensor in nuScenes scans the environment at 20Hz, while the cameras capture images at 12Hz, offering a continuous and reliable stream of data for our tests.

For \tool{} evaluation (Section \ref{sec:adperf_eval}), we also use nuScenes for 3D obstacle detection and trajectory evaluation. For the demonstration, we reuse the above Autoware-AWSIM stack.

\section{Modeling the Performance of ADS}
\label{sec:modeling}

An ADS operates as a real-time system, where each frame must meet strict latency constraints. For instance, at a 10Hz frame rate, any frame exceeding 100ms to process is dropped. Consecutive frame drops lead to a loss of situational awareness, severely compromising driving decisions. However, the complex relationship between detection accuracy and latency under real-world workloads remains unexplored. We are interested in understanding the performance of AI-computation-heavy modules in an ADS and how the performance delays (i.e., latency) would further impact the functionality of an ADS. In this section, we describe how we employ the performance modeling technique \cite{1677534,10.1145/2188286.2188290} to approach this problem.

 

First, we measure detection latency for Apollo on nuScenes and for Autoware on AWSIM with random scenarios to compute the respective service rates.
Second, we model each ADS's obstacle detection component and simulate its performance under varying arrival and computed service rates.
While the measurements reveal the relationship between latency and detection, they do not adequately cover detection behavior in scenarios such as frames with many detections or different sensors with varying arrival rates.
In nuScenes, Apollo detects more than 50 obstacles in less than 0.3\% of nearly 300,000 PCD frames, and in AWSIM random simulation, Autoware detects more than 20 obstacles in less than 0.2\% of 36,000 PCD frames. This scarcity limits our ability to draw comprehensive conclusions about latency in high-density scenarios.
Moreover, different ADS systems utilize diverse sensors with distinct arrival rates, even when using the same detection model. 
Therefore, measuring and modeling the detection module provides valuable insights into the effects of different detection configurations and scenarios, along with their potential adverse impacts.

\subsection{Modeling methodology}
In this subsection, we describe our modeling method for obstacle detection performance in ADSs. Our modeling method consists of four steps. First, we model obstacle detection using a QN~\cite{kleinrock1974queueing, kant1992introduction}. Second, we run a collection of experiments to measure the performance of Apollo and Autoware. Third, we determine the arrival rate ($\lambda$), the rate at which jobs enter a component, and the service rate ($\mu$), the rate at which the component completes jobs, of each queue in the architecture model based on the collected experimental results. Finally, we transform the QNs into a QPN model for simulation.

\pa{Step 1: Modeling the architecture of detection in ADS}

We model the obstacle detection module in ADSs using an open QN with two types of workloads: PCDs from LiDAR sensors and images from cameras. We chose QN to model the ADSs because of their ability to capture multi-component interactions, handle real-world variability, and provide valuable insights into system performance, particularly to identify bottlenecks.
Arrival rate ($\lambda$) 
for obstacle detection is set by the sensors' frame rate. For example, if the LiDAR scans the environment at 20Hz, the $\lambda_{3D}$ for 3D obstacle detection is 20fps.

\begin{figure}[]
\centering
\begin{subfigure}{0.9\columnwidth}
    \includegraphics[width=\textwidth]{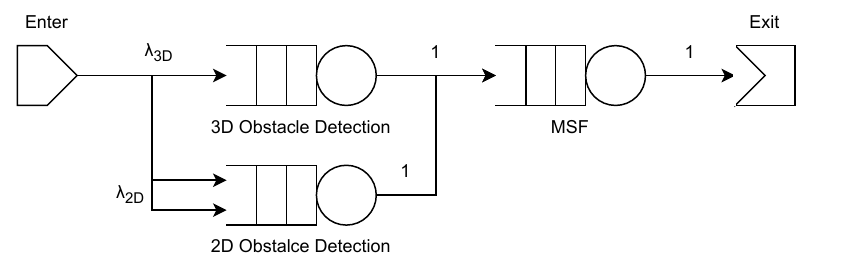}
    \caption{The Queuing Network for Apollo MSF obstacle detection}
    \label{fig:qn_apollo}
\end{subfigure}
\hfill
\begin{subfigure}{0.9\columnwidth}
    \includegraphics[width=\textwidth]{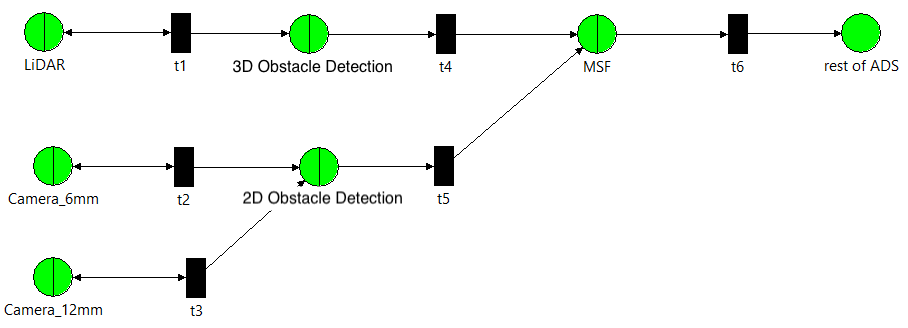}
    \caption{The QPN corresponding to the queueing network}
    \label{fig:qpme_model}
\end{subfigure}
\caption{Performance modeling of Apollo obstacle detection. Modeling figures for Autoware are included in the artifact.}
\label{fig:performance_models}
\end{figure}

Table~\ref{tab:transformation_process} presents the mapping between components from an ADS (Figure~\ref{fig:apollo_sensing_perception}) to a QN model (Figure~\ref{fig:qn_apollo}), then to a QPN model (Figure~\ref{fig:qpme_model}, presented later in step 4). 
We model components as queuing places and their channels as transitions. Each sensor is represented as a source node, resulting in three source nodes and one sink node for the workloads propagated to the rest of the ADS. Since the cameras share the same processing queue, they also share the same type of workload. We model Apollo with three components: 2D detection, 3D detection, and MSF, and Autoware with a single component, 3D detection, as described in Section~\ref{sec:background}. We exclude the radar-based detector as it is used mainly to detect very distant objects outside of the normal scope, i.e., prior works on testing MSF exclude radar-based detectors \cite{yu2022benchmarking, Gao2023BenchmarkingRO, pmlr-v164-tu22a}.

\begin{table}[]
  \caption{Transformation from ADS's Detection to a QPN Model. Only immediate transitions are used.}
  \label{tab:transformation_process}
  \small
  \centering
  \resizebox{\columnwidth}{!}{
\begin{tabular}{lll}
\toprule
Component & QN Element   & QPN Element \\ 
\midrule
Data Channel      & Arc          & Transition   \\
Sensor Input       & Source Node  & Queuing Place, Transition \\
2D/3D Obs. Det., MSF     & Queue        & Queuing Place          \\ 
Rest of ADS        & Sink Node    & Place                  \\
Message            & Workload  & Color                  \\
\bottomrule
\end{tabular}
}
\end{table}

\pa{Step 2: Computing the service rate of each component}

To set the architecture model parameters for performance modeling, we evaluate the live detection performance of Apollo and Autoware, measuring the relationship between the raw number of detections (regardless of IoU) and latency (ms). We run the ADS perception module with live sensor data and record the entry and exit timestamps of each message at every component. Using the native logging frameworks (ROS2 or CyberRT), which capture thousands of messages per run, ensures that the added logging overhead has a negligible impact on latency. Afterward, we calculate the average latency for each component based on the logged timestamps. For live sensor input, Apollo is evaluated using nuScenes dataset recordings (converted for playback), while Autoware is evaluated using AWSIM-generated random scenarios with varying numbers of vehicles.

\textbf{Preparing nuScenes for Apollo replay.} Since nuScenes format is incompatible with Apollo, we use \texttt{adataset}~\cite{adataset} (included in Apollo) to convert nuScenes to Apollo records. 
We map \textit{LIDAR\_TOP} to \textit{$\slash$apollo$\slash$sensor$\slash$lidar128$\slash$compensator$\slash$PointCloud2} and \textit{CAM\_FRONT} to \textit{$\slash$apollo$\slash$sensor$\slash$camera$\slash$\{front\_6mm, front\_12mm\}$\slash$image}. The sensor data incompatible with Apollo's setup is dropped from the recording. 
When replayed in Apollo, the recorded data is fed through mapped channels, emulating real-time sensor input.
\begin{figure}[]
\centering
\begin{subfigure}{0.75\columnwidth}
    \includegraphics[width=\textwidth, trim={0 0 1cm 0},clip]{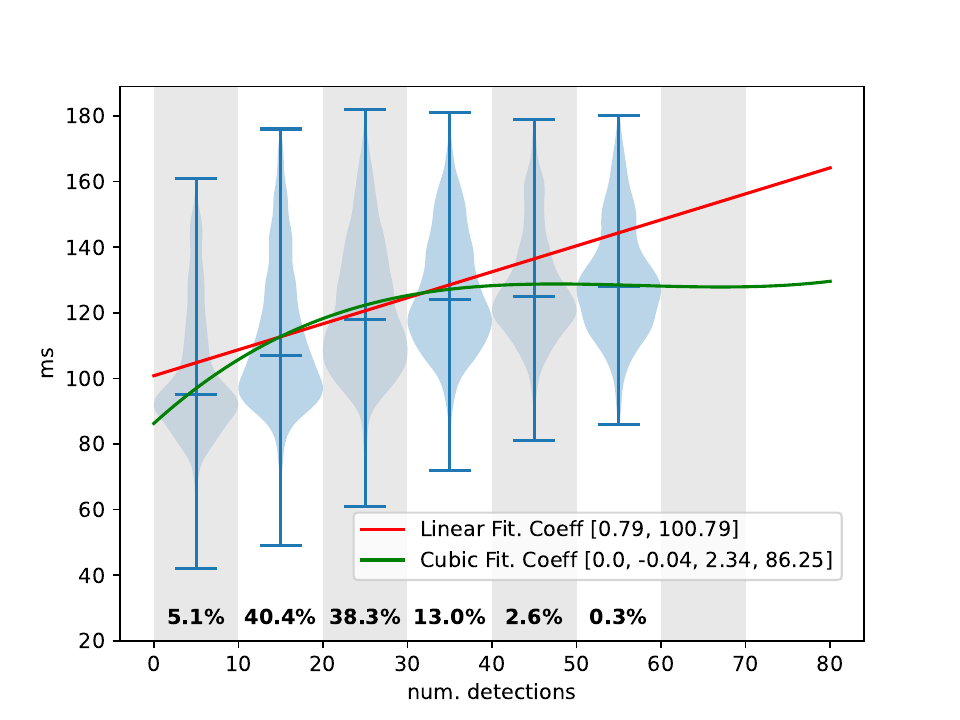}
    \caption{Apollo--nuScenes}
  \label{fig:apollo_lidar_latency_full}
\end{subfigure}

\begin{subfigure}{0.7\columnwidth}
    \includegraphics[width=\textwidth]{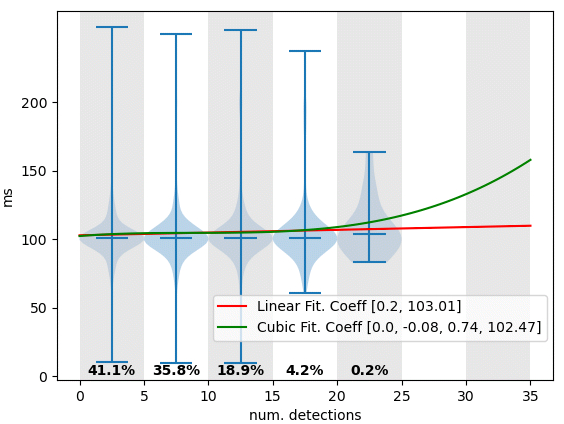}
    \caption{Autoware--AWSIM random scenarios}
  \label{fig:autoware_lidar_latency_full}
\end{subfigure}
\caption{Measured latency of the 3D obstacle detection modules of two ADSs. Latencies are grouped into bins by the number of detections, with the percentage of frames in each bin indicated. Violin plots with marked medians illustrate the latency distribution within each bin. Regression lines reveal a positive correlation between detection count and latency, and are further used to predict latency for unseen frames.}
\label{fig:lidar_latency_full_new}
\end{figure}



\textbf{Measurement Results. } We measure the latency of each available component, i.e., 2D obstacle detection, 3D obstacle detection, and MSF according to Figure~\ref{fig:apollo_sensing_perception}. 
For example, the latency of 3D obstacle detection in Apollo measures the time elapsed between the input from \textit{$\slash$apollo$\slash$sensor$\slash$lidar128$\slash$compensator$\slash$PointCloud2} and the output to \textit{$\slash$apollo$\slash$perception$\slash$innerPrefused}.
Hence, in addition to the three input channels, we monitor an additional two channels: \textit{$\slash$apollo$\slash$perception$\slash$innerPrefused} for both camera and LiDAR output and \textit{$\slash$apollo$\slash$perception$\slash$obstacles} for MSF output to compute latency. In Autoware, as it only performs 3D obstacle detection, we monitor $\slash$perception$\slash$object\_recognition$\slash$detection$\slash$centerpoint$\slash$objects for the detections and lidar\_centerpoint$\slash$debug$\slash$cyclic\_time\_ms for the latency to compute the frequency.

The latency recorded from all three components using a ten-scene sample from nuScenes (Figure included in the artifact) follows an approximately unimodal distribution, with average latencies of 8ms for 2D obstacle detection, 116ms for 3D obstacle detection, and 4ms for MSF, resulting in average service rates ($\mu$) of 125fps for 2D obstacle detection and 250fps for MSF.
This shows that obstacle detection's throughput is limited by the 3D obstacle detection service rate (described in Section~\ref{sec:background}), which is expected to be 20fps based on the nuScenes arrival rate, such 2D obstacle detection and MSF service rate would bear a minimal impact on the performance of the system and is not a performance bottleneck.

On the other hand, 3D obstacle detection shows significantly higher latency, with an average service rate of 8.6fps, much lower than the PCD arrival rate of 20fps. Additionally, 52.7\% of LiDAR frames were dropped during processing, indicating that the 20fps arrival rate exceeds Apollo’s processing capacity.
Further analysis of the 3D obstacle detection across the full dataset confirms the sample-based findings, with a service rate of 8.5fps. The delays are highly concentrated around the median, with between 70.6\% and 79.4\% of frames having latency within 20ms of the median.
As shown by the linear and cubic regression in Figure~\ref{fig:apollo_lidar_latency_full}, there is a positive correlation between the number of detections and latency, suggesting that the latency increases as the number of detections grows.

For the AWSIM-Autoware setup, as Autoware ADS only performs 3D obstacle detection, we measured the average latency of the entire detection pipeline (Figure \ref{fig:autoware_lidar_latency_full}) to be 128.1ms, resulting in the average service rate of 7.8fps where 6.8\% of LiDAR frames are dropped.

\pa{Step 3: Configuring queues in the architecture model}

To conduct a performance simulation with the architecture model that we built from step 1, we need to set the parameters of each component (e.g., queue in the architecture model), including the message processing latency, the workload arrival rate, and the queue size. 
We set up four different configurations reflecting different arrival and service rates. 
 
\textbf{Configuration 1 (Default setup Apollo).} In (1) we configure the arrival rate $\lambda_{2D}$ = 12fps for camera and $\lambda_{3D}$ = 20fps for LiDAR per nuScenes capture rate, while the 2D, 3D obstacle detections and MSF have service rates of $\mu_{2D}$ = 125fps, $\mu_{3D}$ = 8.5fps, and $\mu_{MSF}$ = 250fps respectively as measured by our previous measurement step. 

\textbf{Configuration 2 (Low-workload setup Apollo).} Since $\lambda_{3D}$ = 20fps is double both Apollo and Autoware's default rate, which may result in a bottleneck due to significantly higher arrival rate, in (2), we keep the same service rates but modify the arrival rate $\lambda_{2D}$ = 15fps for camera and $\lambda_{3D}$ = 10fps for LiDAR according to the default rate to investigate how the model performs when the arrival rate is the default rate. 

\textbf{Configuration 3 (Low-latency setup Apollo).} As the service rate for 3D obstacle detection, i.e., $\mu_{3D}$ = 8.5fps, is still lower than $\lambda_{3D}$ = 10fps in the second setup, we are interested in a simulation with a higher service rate for 3D obstacle detection. Thus, we assume that 3D obstacle detection always detects fewer than five detections per the cubic regression shown in Figure~\ref{fig:apollo_lidar_latency_full}, which results in the $\mu_{3D}$ = 10.5fps, where the service rate is greater than the arrival rate.

 
\textbf{Configuration 4 (Low-workload setup Autoware).} Similar to (2), however, $\mu_{3D}$ = 7.8fps to reflect Autoware's measurements.

Since Autoware’s 3D obstacle detection average service rate $\mu_{3D} = 7.8$ fps, and its average service rate for fewer than five detections $\mu_{3D} = 9.7$ fps, are both below 10 fps, we do not report results for configurations (1) and (3) on Autoware, as they would not yield additional insights.

In both Apollo and Autoware, only the newest message is processed while the rest is dropped; this effectively reduces the queue size to one. Therefore, we set the queue size to be one for all setups.


\pa{Step 4: Creating Queueing Petri Nets for simulation}


We use QPME to simulate different load scenarios of detection in Apollo and Autoware, modeled with QPN to find bottlenecks in the setup. The transformation mapping is shown in Table~\ref{tab:transformation_process}.
The resulting QPN model for Apollo, shown in Figure~\ref{fig:qpme_model}, consists of six queueing places and one ordinary place. Each server in the source QN model is mapped to a queueing place (FIFO) in the QPN model. The source nodes are modeled as individual queueing places with the Infinite Server scheduling strategy, initialized with one token.
The service rate for each queueing place follows an exponential distribution, with rates described in step 3. 
The QPN model also includes an ordinary place representing the sink node, which corresponds to the remainder of the ADS. Two colors are used in the model, each representing a distinct workload from the source QN model. The QPN contains six transitions, each corresponding to an arc in the QN model, with each transition having a single mode and a firing weight of one.



\subsection{Simulation results}
During simulation, the following metrics are measured: throughput ($X$), mean service time ($S$), utilization ($U$), and mean token population ($P$). Note that since the queue size of all the queues in the architecture model is one (cf., step 3), any queue length above one would be considered an overflow.

\begin{table}[tbh]
  \caption{QPME simulation results with different configurations: (1) Default setup, (2) Low-workload setup, and (3) Low-latency setup for Apollo and (4) Low-workload setup for Autoware. The following performance metrics are computed: throughput ($X$), mean service time ($S$), utilization ($U$), and mean token population ($P$). (*) shows overflows.}
  \label{tab:qpme_results}
  \centering
  \resizebox{\columnwidth}{!}{
  \small
\begin{tabular}{c|lrrrr}
\toprule
& Metric       & Default       & Low-workload      & Low-latency & Low-workload\\
&              & Apollo        & Apollo            & Apollo      & Autoware  \\
\midrule
\multirow{4}{*}{\rotatebox[origin=c]{45}{2D Obs. Det.}}
& $X$          & 23.99         & 30.004            & 30.009      & NA  \\
& $U$          & 0.192         & 0.24              & 0.24        & NA  \\
& $P$          & 4.038         & 0.316             & 0.316       & NA  \\
& $S$ (ms)     & 0.01          & 0.011             & 0.011        & NA  \\
\midrule
\multirow{4}{*}{\rotatebox[origin=c]{45}{3D Obs. Det.}}
& $X$          & 8.497         & 8.49              & 9.997       & 7.803\\
& $U$          & 1.0           & 1.0               & 0.952       & 1.0\\
& $P$          & 574,828*      & 74,995*           & 20*         & 1,096,202*\\
& $S$ (ms)     & 28,706*       & 7,500*            & 2.004*      & 109,651*\\
\midrule
\multirow{4}{*}{\rotatebox[origin=c]{45}{MSF}}
& $X$          & 32.487        & 38.493            & 40.003     & NA  \\
& $U$          & 0.058         & 0.064             & 0.07       & NA  \\
& $P$          & 0.064         & 0.071             & 0.078      & NA  \\
& $S$ (ms)     & 0.002         & 0.002             & 0.002      & NA  \\
\bottomrule
\end{tabular}
}
\end{table}

The simulation is configured as follows: we set the \textit{Max total run length} to 1,000,000, and the \textit{Simulation stopping criterion} to \textit{Fixed sample size}, meaning the simulation will run until the \textit{Fixed sample size} is exhausted. These parameters are selected to ensure the model is tested with a sufficiently large number of inputs to evaluate throughput and identify bottlenecks.
Table~\ref{tab:qpme_results} shows that 2D obstacle detection and MSF are quite underutilized, and their arrival rates are well below the service rate. 
Table~\ref{tab:qpme_results} presents the results for the configurations (described in step 3). From the results of configuration (1), we observe that 3D obstacle detection is a major bottleneck in the system. Since an ADS processes only the most recent message, any queue with more than one token causes an overflow, resulting in older frames being dropped. As shown, the average number of tokens ($P$) in the 3D obstacle detection queue is consistently greater than one by a significant margin. (Ideally, the simulation would capture the exact number of tokens that overflow when the queue size is set to one, but this functionality is not operational.)
Thus, we experimented with (2). While we observed significant improvements, the bottleneck remains.
Configuration (3) demonstrates the system's performance with very few detections, specifically fewer than five. 
The results in the third column of Table~\ref{tab:qpme_results} show that, although all measurements improve substantially, overflow still occurs.
As LiDAR is the main sensor in Apollo (as described in Section \ref{sec:background}), overflown LiDAR messages result in processed but unfused detections by 2D obstacle detection, reducing the accuracy of MSF perception while wasting computation power for 2D obstacle detection.

For (4), which models Autoware, we observe similar results to (2). As Autoware only uses 3D obstacle detection, it does not represent a detection bottleneck. However, it means that the current servicing rate is not suitable for detecting obstacles at the current input rate, especially in dense scenarios.

\rqboxc{
Our experimental and simulation results show that, for 3D obstacle detection, the number of detections is positively correlated with latency. Since the input rate already exceeds the service rate for both Apollo and Autoware, 3D obstacle detection poses serious challenges in dense scenarios and can become a bottleneck, reducing the throughput of MSF ADSs such as Apollo.}

\section{\tool{}: Automated Performance Tests Generation for ADS}
\label{sec:adf}
Our performance study reveals that bottlenecks in Apollo's obstacle detection are likely to emerge as detections become more diverse and complex, yet datasets (e.g., nuScenes) are not effectively in exposing performance bottlenecks.
Performance modeling (Section~\ref{sec:modeling}) offers theoretical insights and guidance,
but lacks realistic automated tests (i.e., driving scenarios).

To address the mentioned limitations, we propose \tool, the first automated framework for ADS performance testing.
Unlike current accuracy-focused methods, \tool{} generates simple and realistic latency-increasing test scenarios for 3D obstacle detection, the slowest component in obstacle detection.
While multi-sensor synchronization is ideal for MSF-based testing, we focus on LiDAR inputs, as delaying 3D obstacle detection can stall the entire fusion pipeline and make detections from other sensors ineffective.

\pa{Overview of \tool.}
\begin{figure*}[]
  \centering
  \includegraphics[width=0.85\linewidth]{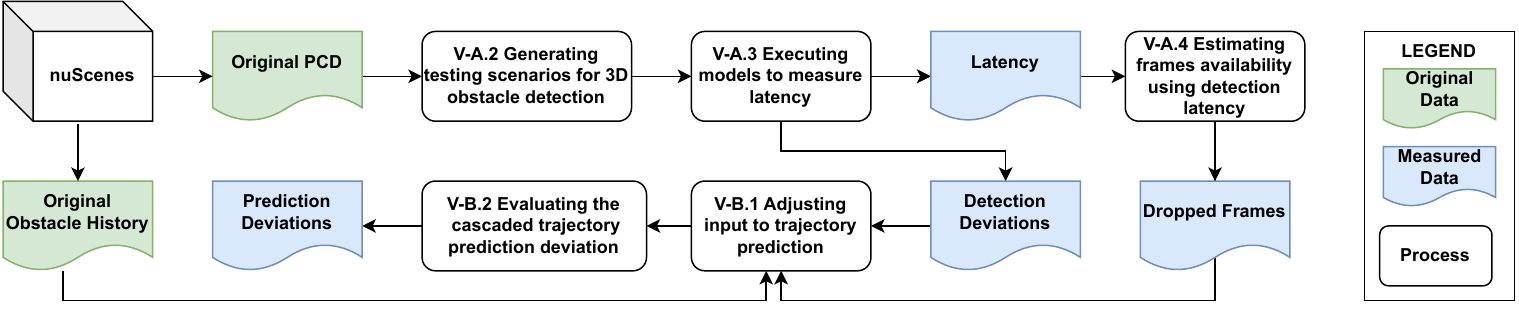}
  \caption{Overview of the \tool, showing how the data flows from the beginning (nuScenes).}
\label{fig:method_approach_overview}
\end{figure*}
Figure~\ref{fig:method_approach_overview} shows an overview of \tool.
Leveraging existing real-world driving scenes (e.g., nuScenes) with ground-truth annotations (GT), \tool{} applies performance-specific mutation operators to each frame in existing scenes, producing new driving scenarios that expose performance issues when evaluated using an ADS.

\tool{} begins by extracting the point cloud representation (original PCD) for each obstacle in every frame of the dataset. It then modifies these obstacles to increase detection rates. After running 3D obstacle detection on the modified PCD frames, \tool{} measures latency, bounding box deviations, and the number of detected obstacles (calculated via intersection-over-union with GT). Using the latency data, \tool{} estimates dropped frames based on ADS frame-dropping rules and evaluates the prediction deviations caused by these drops, assessing their cascading impact on trajectory predictions.

\subsection{Test Generation for 3D Obstacle Detection}
\subsubsection{Data preparation}
\label{sec:method_locate_obstacles}
\begin{figure}[]
\centering
\begin{subfigure}{0.32\columnwidth}
    \includegraphics[width=\textwidth, trim={2cm 6cm 0 6cm}, clip]{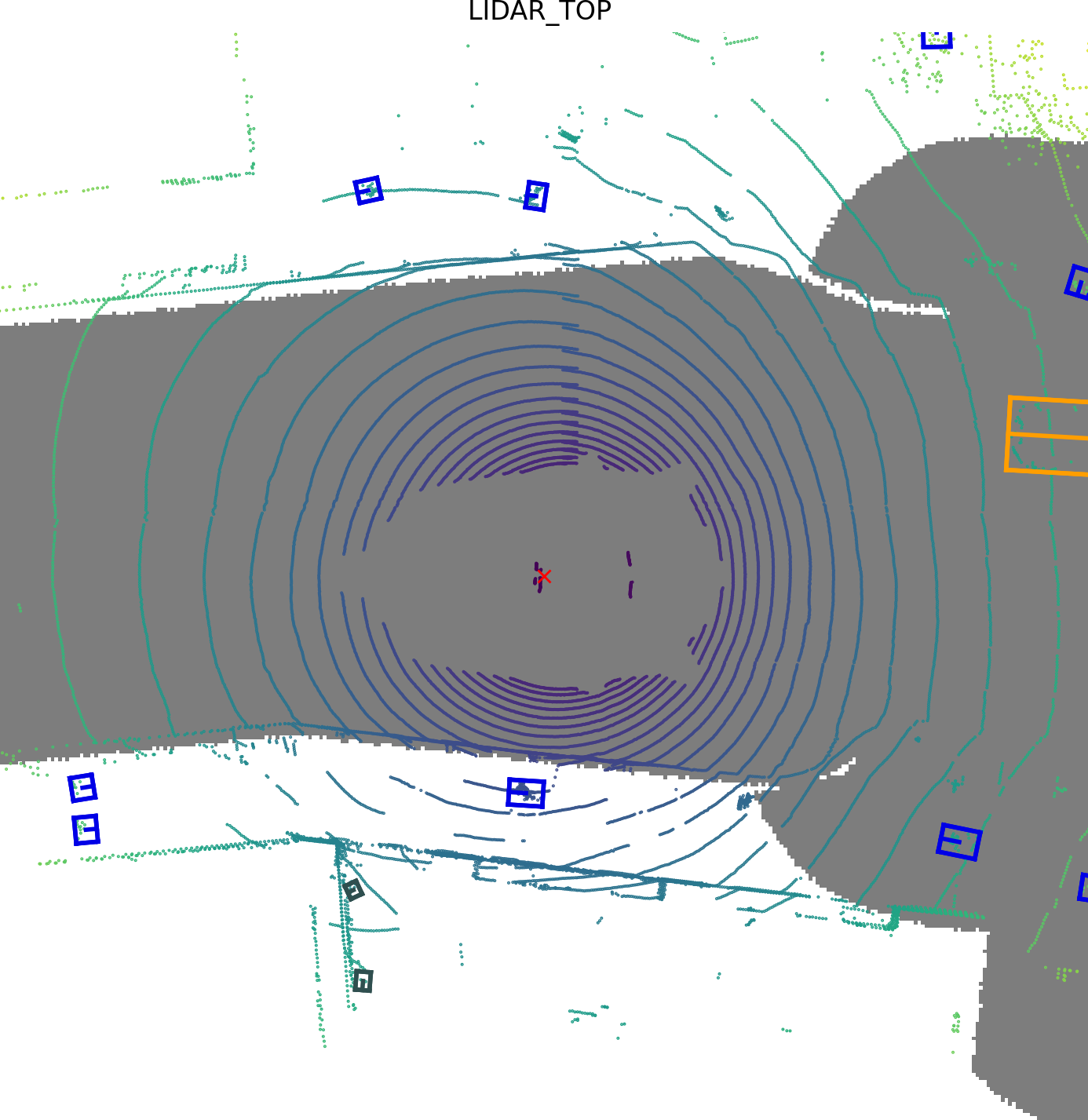}
    \caption{Frame 4}
    \label{fig:trajectory_visualization_4}
\end{subfigure}
\hfill
\begin{subfigure}{0.32\columnwidth}
    \includegraphics[width=\textwidth, trim={2cm 6cm 0 6cm}, clip]{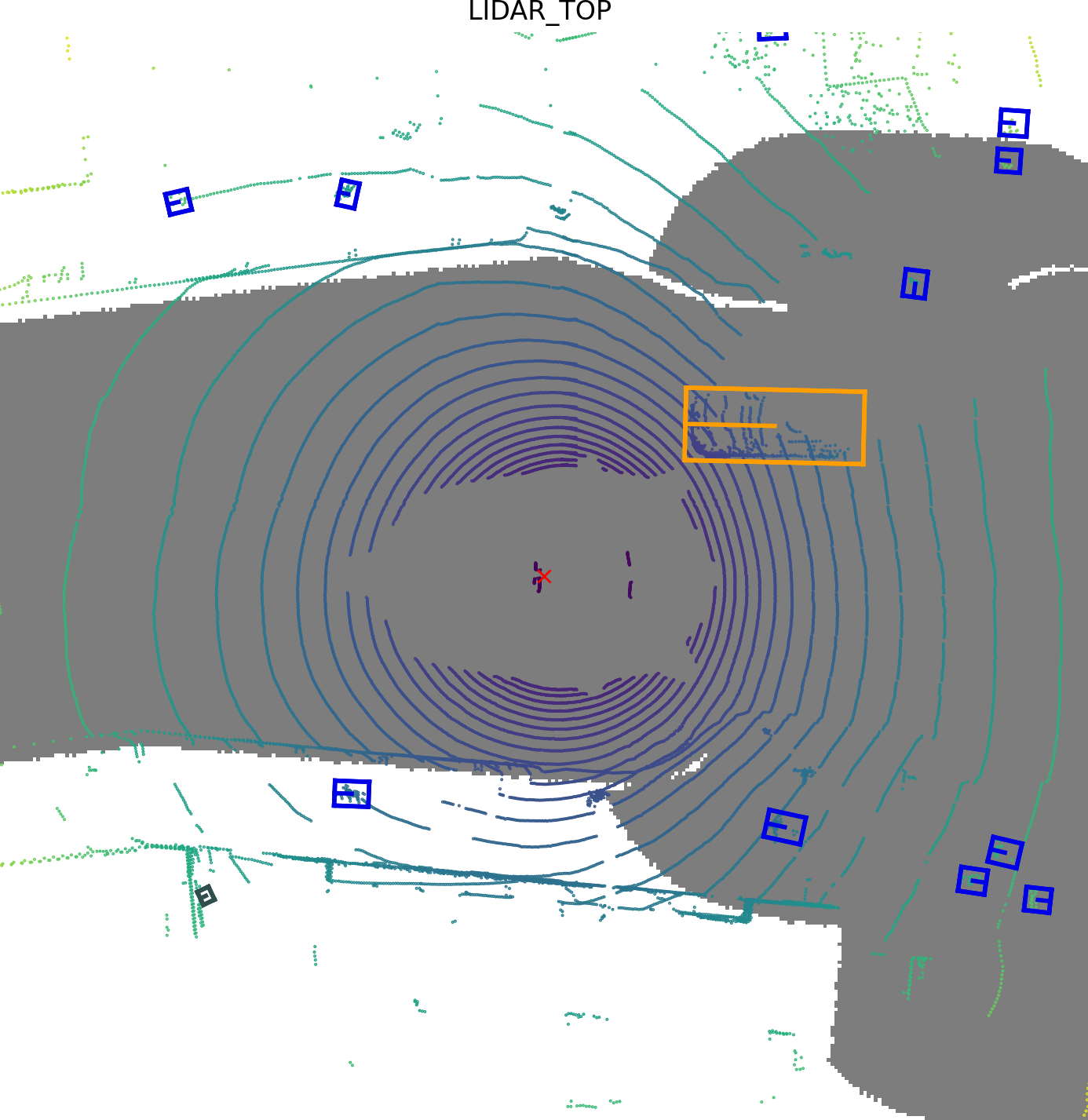}
    \caption{Frame 5}
    \label{fig:trajectory_visualization_5}
\end{subfigure}
\hfill
\begin{subfigure}{0.32\columnwidth}
    \includegraphics[width=\textwidth, trim={2cm 6cm 0 6cm}, clip]{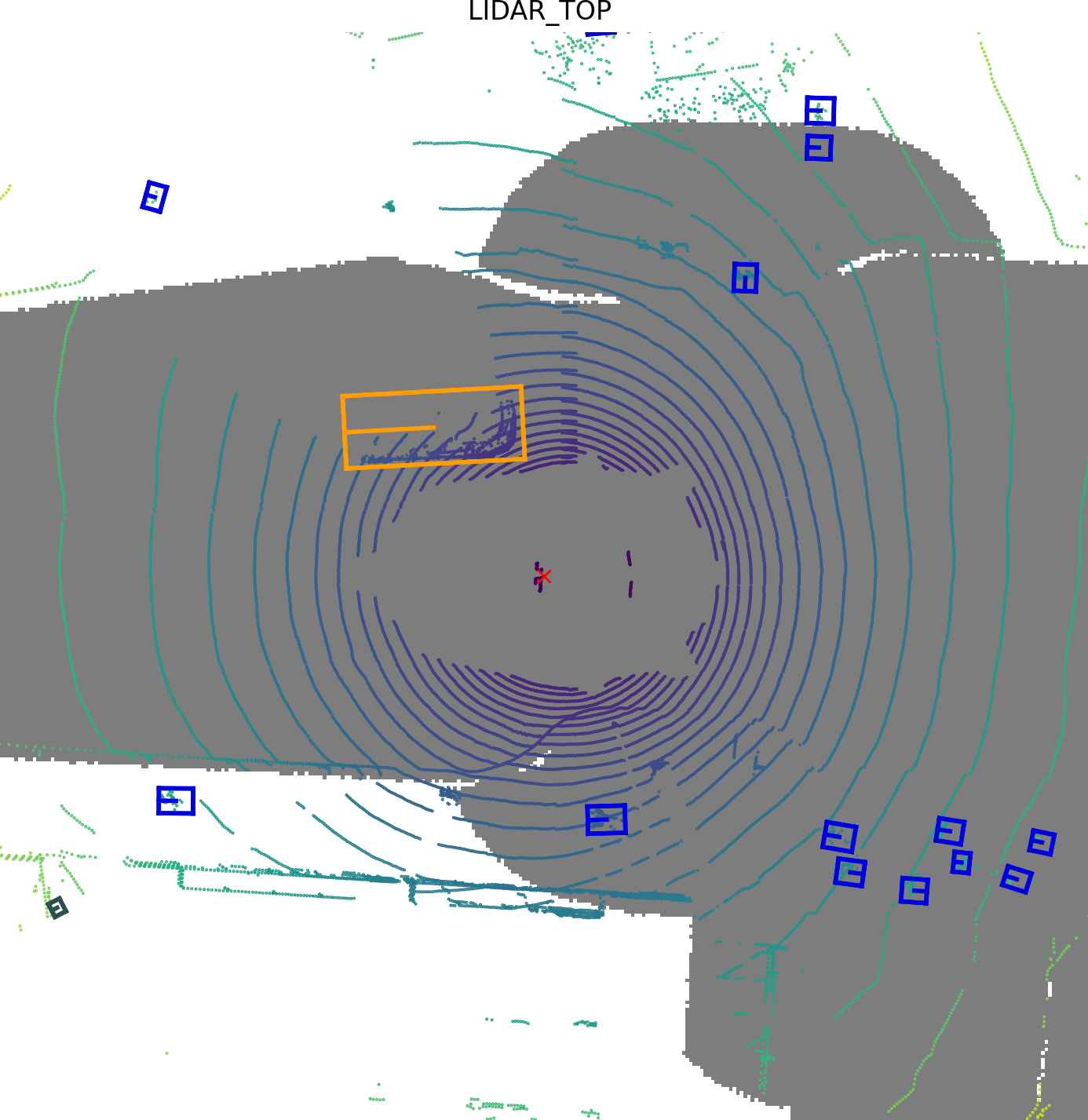}
    \caption{Frame 6}
    \label{fig:trajectory_visualization_6}
\end{subfigure}
\caption{Visualization of three PCD frames showing a moving obstacle car with the orange GT bounding box. The car moves towards the ego car. The size (in point cloud) at each frame is 56, 518, and 698, respectively.}
\label{fig:trajectory_visualization}
\end{figure}
Given an input dataset of real-world driving scenes, \tool{} extracts two types of obstacle data: point cloud representation in PCD and obstacle history data. First, it extracts the PCD scans and the point cloud representations for GT obstacles within the PCD to test 3D obstacle detection. 
In the nuScenes dataset, \tool{} can directly access the point cloud representation of the obstacles through the panoptic dataset to modify the points to generate test scenarios.
It then uses trajectory prediction code to convert obstacle history data (visualized as consecutive frames in Figure \ref{fig:trajectory_visualization}) into a predictor-compatible format.
In obstacle detection, each frame contains all present obstacles (schema A), while in trajectory prediction, each obstacle contains the frames in which it appears (schema B), enabling quick access to each obstacle's history.

\subsubsection{Generating testing scenarios for 3D obstacle detection}
\label{sec:method_test_generation}
In this work, we take a novel approach to robustness test generation by focusing on increasing the number of detections. To achieve this, we modify the LiDAR representations of obstacles based on the test scenario.
Table~\ref{tab:modification_types} outlines the operators used to generate these testing scenarios.

\begin{table}[]
  \caption{\tool{} Tests Operators}
  \label{tab:modification_types}
  \small
  \centering
  \resizebox{0.9\columnwidth}{!}{
\begin{tabular}{l|lll}
    \toprule

Type & Direction        & Distance (m)            & \% of each obstacle \\
    \midrule

Add Noises & $\pm$ y & 0.1, 0.3, 0.5 & 5.8, 17.4, 29    \\
Add Obstacles   & $\pm$ y & 3                       & 100           \\
Move Obstacles  & $\pm$ y                 
                          & 0.1, 0.3, 0.5 & 100      \\        \bottomrule

\end{tabular}
}
\end{table}

\pa{- Adding Noises Outside of the Obstacles Bounding Boxes.}
Sensor data is noisy due to various factors such as environmental conditions, hardware limitations, or interference. In the context of obstacle detection, sensor noise can manifest as random variations and inaccuracies in the sensor readings. 

In contrast to existing work, which adds noises to LiDAR representation of the obstacles (inside of the GT bounding boxes) \cite{yu2022benchmarking, Gao2023BenchmarkingRO, leimalidarrobustnessbenchmark, LiRtest, Dong2023BenchmarkingRO}, we add noises right outside of the GT bounding boxes. These are simple, randomly generated points meant to increase the number of candidate objects detected by 3D obstacle detection by inflating the size of the obstacle. Since filtering these candidates is expensive (up to $O(K^2)$) as described in Section~\ref{sec:related_perf}, increasing the number of candidates has the most impact in increasing the latency while having an overall minimal impact on the detection results \cite{slowlidar, 10650435}. 

We add noise outside each annotated obstacle through three steps: 1) identify noise distance, 2) calculate size, and 3) add noise.
First, we determine the maximum distance ($d_{noise}$) from either the left or right of the obstacles to add noise. We experiment with three distances: 0.1, 0.3, and 0.5m.
Then, we calculate the number of points to add $num_{noise}=num_{obs} \times \frac{d_{noise}}{width_{obs}}$
where $num_{obs}$ represents the count of the point cloud representation of the obstacle and $width_{obs}$ represents the width of the obstacle.
For example, assuming the average obstacle is similar to nuScenes' car (4.08m x 1.73m), then $num\_points_{noise}$ added at 0.1m to the obstacle in Figure \ref{fig:trajectory_visualization_6}, is $698*\frac{0.1m}{1.73m}=40$ points.
As $d_{noise}$ increases, $num_{noise}$ also increases following the formula to ensure that the added noise is not too sparse. 
Finally, \tool{} identifies a region $R_{noise}$ of size $d_{noise} \times length_{obs}$ (where $length_{obs}$ is the length of the obstacle) immediately adjacent to either side of the obstacle.
\tool{} upsample the existing points in the region 
to match $num_{noise}$.
Gaussian noises are commonly used to simulate realistic environment noises~\cite{Dosovitskiy2017CARLAAO,airsim,9967197}. 
Hence, \tool{} perturbs the newly added points with Gaussian noises of 0.05m in line with previous works \cite{yu2022benchmarking, Dong2023BenchmarkingRO}. The noises added ($P_{noise}$) can be summarized by the formula:
\begin{multline}
    P_{noise} = \{ p + normal(0, 1) \times 0.05m \mid\\ p \in upsample(R_{noise}, num_{noise}) \}
\end{multline}
where $upsample(R_{noise}, num_{noise})$ represents the up sampled noise from $R_{noise}$, $normal(0, 1)$ represents Gaussian noises with $\mu=0$ and $\sigma^2=1$, $0.05m$ is the scale of the noise and $p$ represents individual point from the upsampled $R_{noise}$ region.


In nuScenes, we add fewer than 670 points, far fewer than state-of-the-art latency-inducing adversarial methods \cite{slowlidar}.

\pa{- Adding Obstacles.} The next test involves adding new obstacles to the PCD, motivated by the findings in Section \ref{sec:modeling}. 
As the number of obstacles increases, we aim to observe a corresponding rise in latency, confirming their relationship.
These obstacles are selected from the existing obstacles in each frame to ensure realism. The new obstacles are placed two car widths away from the existing ones, for two reasons: First, to minimize the impact of different capture angles on obstacles at varying distances, we avoid placing new obstacles too far away. Second, the new obstacle must account for both the original and its own width while maintaining a small gap.
To simplify the calculations, we assume that the cars have the same width as those in the nuScenes dataset, i.e., 1.73m. As a result, the new obstacles are placed approximately three meters away from the original ones. However, the obstacle can be placed at a different location if needed. The selection and checking process is repeated each frame to maintain plausibility over the entire scene.

\pa{- Moving Obstacles.} The final test involves moving existing obstacles in the current frame. We explore moving the obstacles closer together. The objective is to challenge the detection process by causing the obstacles' point cloud representations to be close but not overlap, thereby confusing the system about the number of obstacles.
To do this, we first compute the center of mass using the LiDAR representation of all obstacles in each frame, then shift them toward the y-center to assess the impact of clustered obstacles on detection latency.

\subsubsection{Executing models to measure latency}
\label{sec:method_run_models}
We run the 3D obstacle detection models on both the unaltered and modified PCD frames for comparison. The models are executed with a single worker, and the data frames are not shuffled to maintain the original ordering. The latency between the model's input and output is measured.

\subsubsection{Estimating frames availability using detection latency}
\label{sec:method_drop_frames}
Since the 3D obstacle detection models are run via OpenPCDet and not within an active ADS (e.g., Apollo), frames are not automatically dropped. As a result, we estimate data availability (i.e., which frames are dropped) based on detection latency. Our research into how existing ADSs handle latency reveals two approaches to managing high latency. In Apollo, incoming messages are stored in a FIFO queue. After the current message is processed, the newest message is selected for processing, while any intermediate messages are considered outdated and dropped \cite{apollo_buffer}. In Autoware, there is no general rule, but the point cloud preprocessor documentation mentions a timer to manage high latency \cite{autoware_timer}. Since this timer is highly specific and requires tuning for each frame rate, we adopt Apollo's approach, as it is more widely applicable. 

Using the sensor message arrival rate ($\lambda_{\text{sensor}}$), \tool{} can compute the expected processing time for each component. For instance, if the sensor message arrival rate is 20Hz (as in nuScenes), the expected processing time for the corresponding component is $\frac{1}{20\,\text{Hz}} = 50\,\text{ms}$. The delay ($delay_{frame_i}$) for each processed message is computed as the latency exceeding the expected processing time. Since the frames arrive at a fixed rate, \tool{} only considers positive delays.
\begin{equation}
    delay_{frame_i}=maximum(0, latency_{frame_i}-\frac{1}{\lambda_{sensor}})
\end{equation}
Using this formula, \tool{} computes the delays for all frames ($delay\_by\_frame$). Then, the availability for each frame can be estimated based on the accumulated delay (Algorithm~\ref{algo:frame_drop}).

\begin{algorithm}
    \caption{\tool's Frame Drop Computation \label{algo:frame_drop}}
    \KwData{$delay\_by\_frame$, $theshold$}
    \KwResult{indices of points dropped}
    $accm\_delay \gets 0$\, $dropped\_frames \gets Empty\ list$\;
    \For{each $frame\_index$, $delay$ in $delay\_by\_frame$}{
        \If {$frame\_index$ is start of new scene}{
            \tcc{restart for each scene}
            $accm\_delay \gets 0$\;
        }
        \If {$accm\_delay \geq threshold$}{
            \tcc{adjust acc. delay and save alarm}
            $accm\_delay \gets Max(0, accm\_delay - threshold)$\;
            Append $frame\_index$ to $dropped\_frames$\;
        }
        \Else{
            $accm\_delay \gets accm\_delay + delay$\;
        }
    }
\end{algorithm}

\subsection{Test Generation for Trajectory Prediction}
\subsubsection{Adjusting input to trajectory prediction}
\label{sec:prepare_trajectory_prediction}
We aim to assess the impact of dropped frames from 3D obstacle detection on trajectory prediction. Before doing so, we must adjust the GT to reflect the PCD modifications-induced changes. Without this adjustment, two issues arise: first, the frame drop may not occur without the PCD modification, and second, deviations may be inaccurate if obstacles are moved but not accounted for. For instance, if the deviation is reported as 2m, but we moved the obstacle by 2m, there would be no actual deviation.

\pa{Calculating the deviations from obstacle detection and matching them to the correct GT obstacles.}
For each modified frame, we record latency, detections, and positions to compute the new IoUs and GT matches.
The detected obstacles are matched to the trajectory prediction input in two steps.
First, for each frame, \tool{} adjusts the GT obstacles based on the generated test. For adding noise, the original GT remains unchanged. For moving obstacles, the GT is shifted according to the corresponding changes in the PCD. Added obstacles are ignored, as they do not exist in the GT. 
This enables \tool{} to match the detected obstacles from the modified PCD to the modified GT obstacles using IoU, allowing the computation and linkage of detection deviations to the corresponding GT. 
We consider two types of deviations, i.e., status and displacement.
For status deviations, we only consider when GT obstacles are \textbf{not} detected. 
When GT obstacles are detected, we consider their displacement.
These deviations are used to adjust the trajectory prediction input.

\pa{Adjusting the input of trajectory prediction using the calculated deviations.}
To transfer the observed deviations from the previous step, we first match the obstacles in the two different data schemas using GT and frame ID.
For each deviation type described, we handle the obstacles differently.
If the histories are too short or fragmented, trajectory calculations would either be impossible or highly inaccurate. Repeating the status strikes a balance between maintaining calculable and reasonably accurate obstacle trajectories.
For displacements, \tool{} shifts the obstacles in the trajectory prediction input according to the displacement values.

\subsubsection{Evaluating the cascaded trajectory prediction deviation}
After adjusting the previous step, we can proceed with frame dropping and estimate the impact. Since we use datasets containing scenes with consecutive data frames, we can drop the exact frames where data is expected to be dropped based on obstacle detection latency. Following Apollo's rule, we calculate the frames to drop as outlined in Subsection~\ref{sec:method_drop_frames}.
To drop these frames from the trajectory prediction input, we iterate over each obstacle and each frame.
If the index of the frame matches a dropped frame, we set its values to those of its previous frame. If all frames of an obstacle are dropped, we set its values in all frames to be those of the first frame, as if the obstacle is stationary throughout the scene.
Once we run trajectory prediction with the corresponding frames dropped from the calculation, we obtain a list of obstacles and their predicted trajectory. Comparing the newly generated trajectory with that when no frames are dropped, we can obtain the deviation to analyze the impact of increased latency.

\section{Evaluating ADS Performance Using \tool}
\label{sec:adperf_eval}
This evaluation contains three experiments. First, we evaluated the robustness of two 3D obstacle detection models, i.e., PointPillar and CenterPoint, against the latency-stressing scenarios by \tool. Second, we investigate how the high-latency obstacle detection may affect trajectory prediction. Finally, we demonstrated in simulation that increased 3D obstacle detection latency cascades through the ADS, leading to adverse outcomes such as indefinite stops.

\subsection{Testing 3D obstacle detection availability robustness to latency increasing scenarios}
\pa{Motivation.}
From the previous experiment, we identify 3D obstacle detection as a performance bottleneck in ADS, with latency increasing as the number of obstacles grows. This motivates an investigation into its resilience to noise and variations, as its robustness directly impacts system throughput and real-world reliability.

\pa{Method.} We assess the robustness of 3D obstacle detection in ADSs to noise and changes while measuring the sensitivity of latency through three simple and realistic tests: adding noise, adding obstacles, and moving obstacles in the PCD from nuScenes. These mutations are carried out following the procedures outlined in subsections \ref{sec:method_locate_obstacles}, \ref{sec:method_test_generation}, \ref{sec:method_run_models}, and \ref{sec:method_drop_frames}.
For the 3D obstacle detection, we select two models from the OpenPCDet model zoo~\cite{openpcdet2020}, both trained on nuScenes data: PointPillar-MultiHead (PointPillar) and CenterPoint-PointPillar (CenterPoint). These models were chosen for their role in well-known ADSs: PointPillar in Apollo and CenterPoint in Autoware.

To evaluate whether the observed increased latencies with \tool's tests are statistically significant, we apply non-parametric methods that account for the distribution of the data. We applied both Cliff's delta ($\delta$) \cite{romano2006} and Wilcoxon signed-rank test~\cite{Wilcoxon1945} to compare the two sets of latency values, i.e., the original latency values and the ones under each perturbation type. 

\pa{Results.}
\begin{figure}[]
  \centering
  \includegraphics[trim={0 0.05cm 0 0},clip,width=0.8\linewidth]{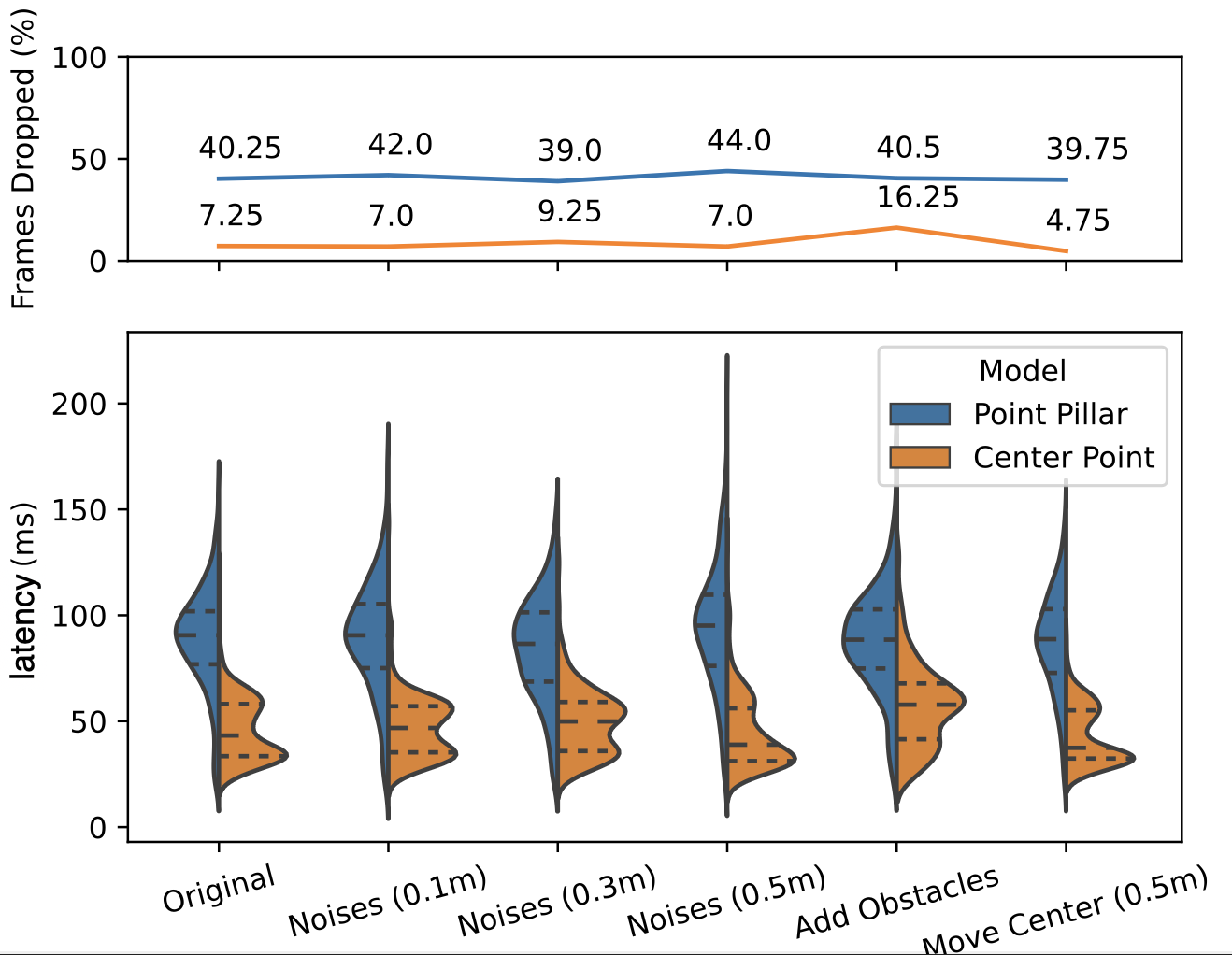}
  \caption{Effects of different tests on model latency distributions. The violin plots show quartiles (dashed lines), and the linear plot indicates the percentage of frames dropped.}
  \label{fig:rq2_latency_results}
\end{figure}
Overall, PointPillar generates 84 fewer raw detections on average than CenterPoint for the same number of detected obstacles. Despite this, PointPillar has significantly higher latency, with the average median latency 41ms higher and the maximum latency 43ms higher, as shown in Figure~\ref{fig:rq2_latency_results}. The linear graph also shows that obstacle detection using PointPillar causes considerably more frame drops, where the median latency is 88-95ms. Additionally, PointPillar shows a smaller increase in detections from the tests. However, the increase in detections has a more significant impact on latency.

Both models exhibit latency sensitivity to the generated testing scenarios, albeit in different ways.
PointPillar is more sensitive to added noise (mean latency increases up to 6.2ms and the frame drop rate increases up to 3.75\%), while it remains relatively resistant to added obstacles (latency decreases by 0.4ms and frame drop rate increases 0.25\%). 
In contrast, added noises increase the average latency up to 3.9ms and the number of frames dropped up to 2\% in CenterPoint, while added obstacles increase the mean latency by 11.5ms and the percentage of dropped frames to 16.25\%. 
Both models exhibit strong resistance to moving obstacles, with the frame drop rate decreasing by 4.25\% for PointPillar and 3.5\% for CenterPoint.

\pa{Statistical Analysis.} For PointPillars,  two types ``noise (0.3m)'' and ``noise (0.5m)" show statistically significant increased latencies ($p-value$ is  0.02 and $< 0.001$ respectively) with small effect size ($r = 0.12, 0.20$ respectively). 
Cliff's delta confirmed the small effect for noise (0.5m) ($|\delta|=0.15$), while indicating negligible effects for the remaining PointPillars tests.

For CenterPoint, based on the Wilcoxon test, three types, i.e., ``noise (0.3m)'', ``noise (0.5m)'', and ``moving center (0.5m)'' led to significant latency increases ($p < 0.001$, $p = 0.03$, and $p < 0.001$, respectively), with a small effect size ($r = 0.19$, $r = 0.12$, $r = 0.15$). ``adding obstacles'' produced a highly significant increase ($p < 0.001$) with a medium effect size ($r = 0.48$). 
Only ``noise (0.1m)'' failed to produce a significant difference. This is also supported by Cliff's delta: except for noise at (0.1m), which showed negligible effects, noise (0.3m), noise (0.5m), and moving center (0.5m) yielded small effects ($|\delta|=0.16$, $|\delta|=0.15$, $|\delta|=0.16$), while adding obstacles produced a medium effect ($|\delta|=0.35$).


This experiment reveals contrasting model behaviors under different scenarios. PointPillar is more sensitive to 2 of 3 noise types but handles crowded scenes better, while CenterPoint is less affected by noise but more sensitive to added obstacles.

\rqboxc{
Each obstacle detector has its inherent latency, which \tool's testing scenarios can further amplify. \tool{} triggered opposite responses from the detectors, with each being sensitive to different scenarios. Depending on the specific scenarios the system is often used in, the appropriate detector should be selected for optimal performance.
}

\subsection{Testing trajectory prediction robustness to unavailable detection output}
\pa{Motivation.} Having learned that certain modifications can impact latency, we are eager to explore how these changes affect trajectory prediction, which relies on obstacle detection as its input. Understanding this relationship is crucial, as the unavailability of detection output can significantly influence the accuracy and reliability of trajectory predictions, ultimately shaping the effectiveness of the entire ADS.

\pa{Method.} The changes observed in perception are two-fold: altered detections and their associated latency can influence the prediction results. As a result, we adjust the input as described in Subsection \ref{sec:prepare_trajectory_prediction}.
We use Trajectron++ \cite{salzmann2021trajectron}, a well-established trajectory prediction, which is commonly evaluated in prior works~\cite{Zhang_2022_CVPR, Ivanovic2022ExpandingTD, Gu2022StochasticTP}. 
We measure trajectory deviation using Average Displacement Error (ADE) and Final Displacement Error (FDE), commonly used in prior works~\cite{salzmann2021trajectron, Zhang_2022_CVPR}. ADE reflects overall accuracy, while FDE measures the precision of the final position.
Last, we applied statistical tests to highlight statistical significances when comparing the two metric values between the original and under perturbations by \tool.

\pa{Results.}
\begin{figure}[]
  \centering
  \includegraphics[width=0.8\linewidth]{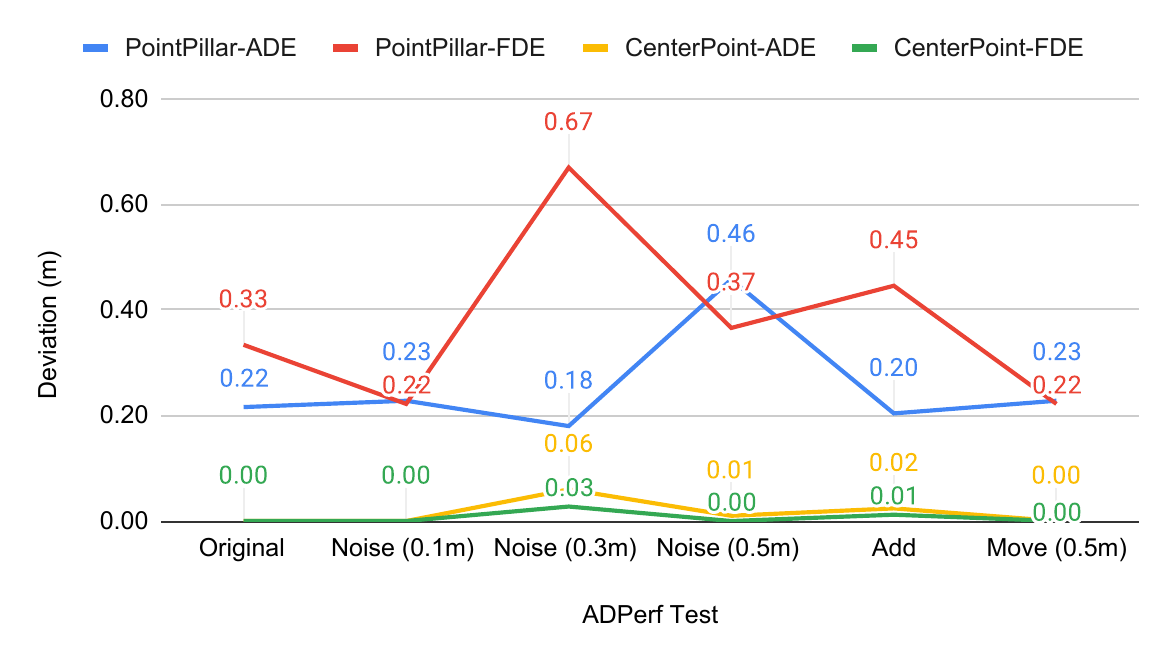}
  \caption{Average ADE and FDE due to data unavailability caused by \tool's tests (Figure~\ref{fig:rq2_latency_results}). Only ADE and FDE of moving obstacles are considered.}
  \label{fig:rq3_deviation_summary}
\end{figure}
Figure \ref{fig:rq3_deviation_summary} shows that despite high frame drop rates, trajectory deviations remain small, indicating Trajectron++'s resilience to missing data, though some trends still emerge.

Overall, the trajectory prediction deviations using CenterPoint’s detection output are significantly smaller than those with PointPillar. Added noise contributes to the largest increase in ADE (0.06m) and FDE (0.03m). With fewer frames dropped from CenterPoint detection (indicating higher availability), trajectory prediction using its output is more resistant to deviations.
On the other hand, the trajectory deviations resulting from PointPillar's detection are significantly larger. Adding noises increases the ADE up to 0.24m and FDE up to 0.34m. Additionally, adding obstacles leads to a 0.12m increase in FDE.
The limited effect on trajectory prediction can be partly attributed to the availability distribution of detection frames. Even under high drop rates, at most two consecutive frames are lost. At 10fps, sufficient temporal information remains available for the prediction module, minimizing the overall impact on trajectory estimation.

Our statistical tests show that the prediction trajectory deviations caused by \textit{CenterPoint}'s increased latencies are not statistically significant. Differently, for \textit{PointPillar}, under all the perturbation types, except for ``noise (0.5m)'', ADE and FDE are statistically significant ($p < 0.02$) with large effect sizes ($0.83 < r < 0.87$). Such statistical significances are also supported by Cliff's test.

Deviations to predicted trajectories can result in unsafe trajectory plans or misleading control decisions, causing path deviation. Such system failures are ultimately linked to real-world collisions as documented in public reports \cite{Tang2022ASO}.

\rqboxc{Trajectory prediction demonstrates resilience to low data availability. However, the availability of data from obstacle detection significantly impacts deviations. Therefore, trajectory prediction should prioritize detectors with higher data availability.}

\subsection{System demonstration: applying \tool{} to test ADSs in simulated environments}
\begin{figure}[]
    \centering
    \includegraphics[width=0.7\linewidth]{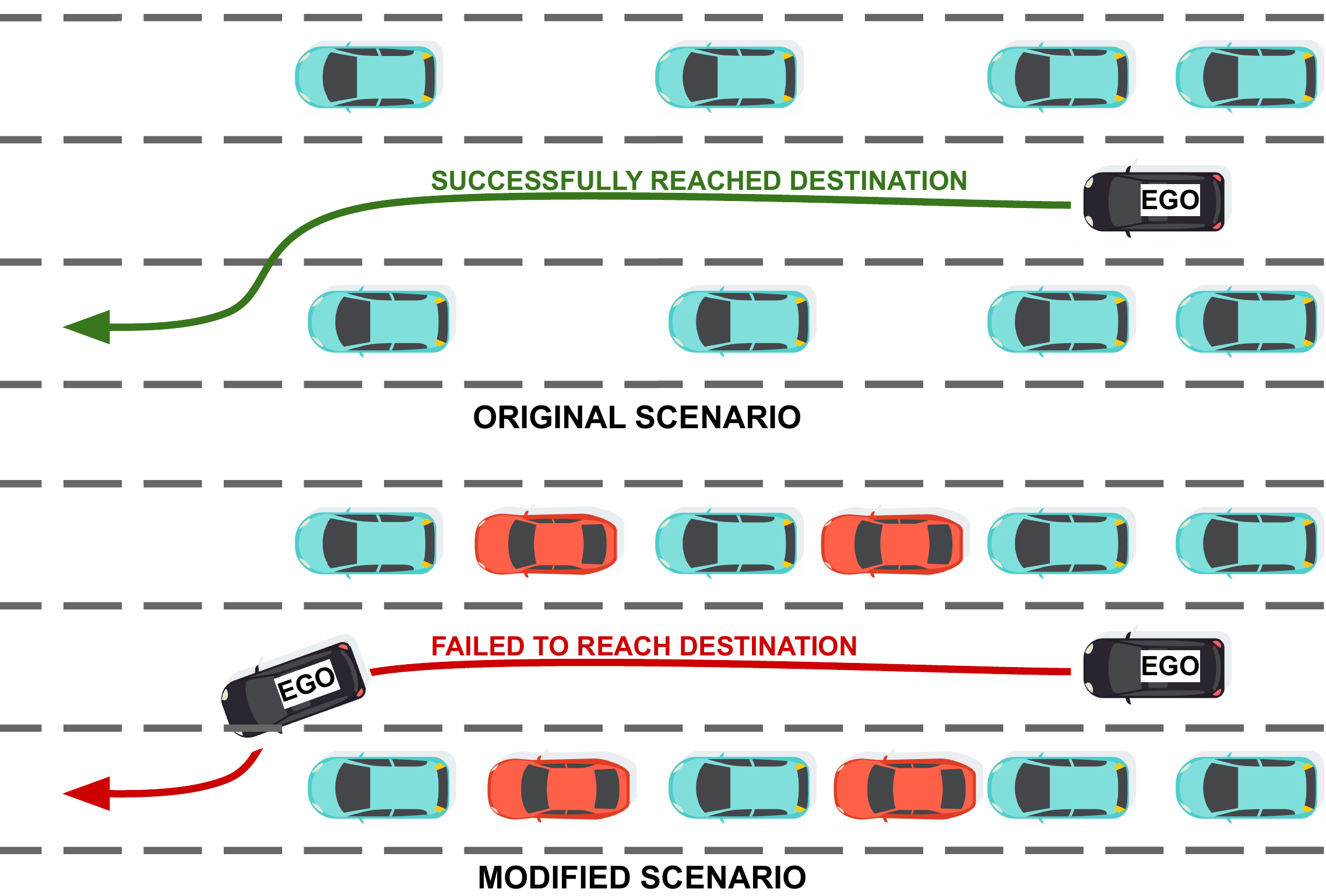}
  \caption{An example scenario in the system demonstration, i.e., exercising Autoware in AWSIM using one latency-stressing scenario by \tool. \tool{} generates a new scenario by adding the red vehicles.}
  \label{fig:demo_scenario}
\end{figure}



\pa{Motivation.} Prior studies \cite{Haq2021CanOT, stocco2023modelvssystem} illustrate that non-trivial gaps exist between testing individual modules alone and testing the whole ADS (e.g., through a simulator). Hence, we set off to demonstrate the impact of increased latency of 3D obstacle detection latency detected by \tool{} on an ADS in a simulated environment.

\pa{Method.} We conduct system demonstrations on Autoware SS2\cite{autoware_robotec} using a compatible simulator (i.e., AWSIM V1.2.0\cite{awsim}). 
We cannot experiment on Apollo due to the lack of a compatible simulator.
AWSIM supports a very limited driving scenario compared to nuScenes, which was used in the other evaluations of this work.
We experimented with different scenarios (e.g., mutating obstacles) but found similar outcomes. In the paper, we focus on one representative demo as an example: In the original scenario (Figure~\ref{fig:demo_scenario}), the ego car needs to travel 60m, along the lane to its destination on the adjacent lane, which means the ego car needs to plan a route for lane changing at a proper time.  Eight vehicles are on the adjacent lanes moving in the same direction. 
The ego car successfully arrived at the destination in 30 seconds.

We then applied \tool{} to generate new scenarios that stress the latency of 3D obstacle detection, and further impact the whole ADS. Limited by space, we will describe one type of perturbation by \tool, which is ``add obstacle''. 

\pa{Results.}
Under the \tool's generated scenarios (i.e., with four non-overlapping vehicles on the adjacent lane), the ego car stops indefinitely and fails to change the lane when it is safe (the bottom figure in Figure~\ref{fig:demo_scenario}).

\underline{On the surface}, we found that an increased number of high-latency (i.e., over 100ms) single-frame processing is observed: from 13.8\% to 23\% of all the frames. High-latency frames are dropped by a real-time AD system. Hence, this translates to a sharp increase in frame dropping. Even though the accuracy of the obstacle detection is consistent between the original and \tool's generated scenarios. The \tool's generated scenario stresses the latency of the obstacle detection, promotes more frame dropping, and causes the ego car to ``freeze.''

\underline{Root cause analysis.} 
We further examine the full story behind the ego car's erroneous behavior. We found that due to the increased frame dropping, some obstacles of the dropped frames are deemed as \textit{disappeared} by the \textit{tracking} module. An obstacle must be continuously (i.e., over one second) not detected to be deemed as ``disappeared''. That exact obstacle in the next frame will have a brand-new status for trajectory prediction, i.e., no history locations, which leads to a poor predicted trajectory. Significant discrepancy between a trajectory and the detected location will lead to frequent replanning, which is computationally expensive. Therefore, the ego car will ``freeze'' once it begins frequent replanning.

\section{Threats to Validity}
\label{sec:threats}
\pa{Experiment reproducibility.} As live performance measuring can be fickled due to randomization of the algorithms pre-processing data for the detection models, for each PCD modification, we evaluate it with the model five times to record the combined latency and enhance reproducibility.



\pa{Scenes contain a large percentage of stopped cars.}
The scenes used to test trajectory prediction contain a low percentage of moving cars (18.5\%), which skews the results by reducing the number of deviations and limiting the variability in the computed deviations. We plan to address this limitation in future studies.




\section{Conclusion}
\label{sec:conclusion}
In this paper, we present a study to investigate the performance of two ADSs that identify the bottleneck in 3D obstacle detection that can be further tested for robustness. We introduce \tool{}, a tool that generates testing scenarios to assess the robustness of 3D obstacle detection and its cascading effects on other components in the ADS, such as trajectory prediction. Experimental results show that the scenarios generated by \tool{} can cause significant delays in 3D obstacle detection, leading to non-trivial deviations in the entire ADS. 
The generated QPME model and code for ADPerf are available at \url{https://github.com/anonfolders/adperf}.


\bibliographystyle{IEEEtran}
\bibliography{main}







\end{document}